\documentclass[12pt]{article}

\usepackage{a4wide}
\usepackage[pdftex,usenames,dvipsnames]{color}
\usepackage{graphics}
\usepackage{amsfonts}
\usepackage{amssymb}

\newcommand{\nit}{\noindent}

\newcommand{\np}{\newpage}

\newcommand{\vs}[1]{\vspace{#1 ex}}
\newcommand{\hs}[1]{\hspace{#1 em}}
\newcommand{\bfr}{\begin{flushright}}
\newcommand{\efr}{\end{flushright}}
\newcommand{\bc}{\begin{center}}
\newcommand{\ec}{\end{center}}
\newcommand{\ben}{\begin{enumerate}}
\newcommand{\een}{\end{enumerate}}

\newcommand{\be}{\begin{equation}}
\newcommand{\ee}{\end{equation}}
\newcommand{\ba}{\begin{array}}
\newcommand{\ea}{\end{array}}
\newcommand{\ct}{\cite}
\newcommand{\bit}{\bibitem}

\newcommand{\bg}{\beta}
\newcommand{\gam}{\gamma}

\newcommand{\kg}{\kappa}

\newcommand{\sg}{\sigma}

\newcommand{\Del}{\Delta}

\newcommand{\Lb}{\Lambda}

\newcommand{\bff}{\bold{f}}

\newcommand{\bfer}{\bold{r}}

\newcommand{\bfv}{\bold{v}}
\newcommand{\bfx}{\bold{x}}

\newcommand{\lh}{\left(}
\newcommand{\rh}{\right)}
\newcommand{\ld}{\left.}

\newcommand{\nb}{\nabla}

\newcommand{\der}{\partial}

\begin{document}

\bc
{\bf \Large  The path to Special Relativity }
\vs{4}

{\bf \large Gideon Koekoek} \\
 \vs{2}
 
Dept.\ of Grav.\ Waves and Fund.\ Physics \\
\vs{1}

Maastricht University, Maastricht \\
\vs{1} 

and Nikhef, Amsterdam\\ 
\vs{3}

{\bf \large Jan W.\ van Holten} \\
\vs{2}

Nikhef, Amsterdam \\
\vs{1} 

and Lorentz Institute, Leiden University, Leiden \\
\vs{3}

{\bf \large Urs Wyder}\\
\vs{1}

Detection and Measurement, Applied Physics Dept. \\
\vs{1}

Fontys University of Applied Science, Eindhoven
\\
\vs{3}

Sept.\ 20, 2021
\ec
\vs{5}

\nit
{\small {\bf Abstract} \\
Following an early observation of Ignatowsky, we present a derivation of the 
transformation rules between inertial systems making no other assumptions 
than the existence of the latter, and show that generically these rules are 
characterized by a constant of nature with the dimensions of an inverse velocity 
squared having the same value in all inertial frames. No independent postulate 
of the existence of an absolute velocity for light waves or other carriers of 
physical information is necessary. Aside from the usual Lorentz- and galilean 
transformations, our analysis also allows for the existence of four-dimensional 
Euclidian transformation rules between inertial systems.
}

\np
\section{Special Relativity \label{s1}}

In his revolutionary article of 1905 on the {\em Electrodynamics of moving 
bodies} Einstein derived the theory of Special Relativity, in particular the 
Lorentz transformations, on the basis of two assumptions; the first assumption
was the existence of inertial frames in which the laws of mechanics and 
electrodynamics always take the same form; and the second one was the 
postulate that the speed of light in vacuum, $c$, is the same in all these 
frames, independent of any relative motion\ct{einstein:1905, einstein:1956}. 

However, the second postulate is superfluous: special relativity holds independent 
of whether light propagates with the same speed in all inertial frames. This was 
already noticed in 1910 by Ignatowsky \ct{ignatowsky:1910}. Indeed, although 
historically the invariance of the speed of light as observed in different inertial 
frames led to the discovery of the Lorentz transformations 
\ct{lorentz:1895,larmor:1898,lorentz:1904}, from a modern perspective a 
spontaneous breaking of gauge invariance could have made photons massive 
without spoiling special relativity. At the time the fact that the speed of light does 
not reveal motion with respect to an electromagnetic aether, and that the Lorentz 
transformations imply that any such aether effects are unobservable, was 
considered of utmost importance. In hindsight it could have been otherwise, 
and electrodynamics is only a particular instance of a relativistic theory of fields 
and particles. Many other phenomena, from time dilation to gravitational waves 
or the behaviour of quarks and leptons with their color and weak interactions, 
are observed confirming special relativity as the theory of relations between 
inertial frames. Therefore it is instructive to reconsider the argument of 
Ignatowsky and discuss special relativity without taking recourse to Maxwell's 
theory and the propagation of light.

In the following we show, that the mere assumption of the {\em existence} of 
inertial frames suffices to prove that there exists a constant of nature $a$ (by 
necessity taking the same value in all inertial systems) from which the existence 
of a universal invariant speed $c$ in all inertial frames can be inferred. This 
invariant speed does not \emph{a priori} need to be interpreted as the speed 
of light, although it is an experimental fact that light does propagate at this speed. 
The nature of the relation between inertial frames is determined exclusively by 
the value of $a$, which can be zero, positive or negative, giving rise to either 
galilean transformations, Lorentz transformations or 4-dimensional euclidean 
rotations.

\section{Inertial frames \label{s2}}

In order to stress the generic nature of transformations between inertial reference 
frames to be derived, we will proceed by using a purely kinematical definition of 
an inertial frame: an inertial frame is characterized by the absence of fields of 
force: electromagnetic, gravitational or otherwise, and the property that all 
free non-interacting particles in such frames move uniformly on straight lines. 

Such frames are not unique; any reference frame related to an inertial frame 
by a constant translation or rotation is an inertial frame as well. A more 
complicated question is, what the relative state of motion of inertial frames 
is allowed to be. This question we will answer first. 

Let $I$ and $I'$ be two inertial frames with associated time- and space co-ordinates. 
The co-ordinates in each frame being unique labels of space-time points, they must 
be related by well-defined functions 
\be
t' = f^0(t,x,y,z), \hs{1} x' = f^1(t,x,y,z), \hs{1} y' = f^2(t,x,y,z), \hs{1} z' = f^3(t,x,y,z).
\label{e2.1}
\ee
Now let $\bfer_p(t) = \bfer_p(0) + {\bf w}_p\, t$ describe the path of a free particle $p$ 
in system $I$; the path of the particle with respect to the reference frame $I'$ then 
is $\bfer'(t')$ such that it moves with a velocity 
\be
{\bf w}_p'(t') = \frac{d\bfer'_p}{dt'} = 
 \ld \frac{\der_t \bff + {\bf w}_p \cdot \nb \bff}{\der_t f^0 + {\bf w}_p \cdot \nb f^0} \right|_p.
\label{e2.2}
\ee
By definition the frame $I'$ is an inertial frame if this velocity is constant for 
{\em all} and any possible particles $p$ with any constant velocities ${\bf w}_p$. 
This implies that the gradients of the transformation functions (\ref{e2.1}) must 
be constant: $\der_t f^a =$ constant, $\nb f^a =$ constant, for $a = (0,1,2,3)$. 
Therefore arbitrary inertial frames are related (in their region of overlap) by 
linear transformations between space-time co-ordinates $x^a = (t,x,y,z)$ and 
$x^{\prime a} = (t', x', y', z')$: 
\be
x^{\prime a} = x^a_0 + \Lb^a_{\;\,b} x^b. 
\label{e2.3}
\ee
In particular this confirms that inertial frames can move with respect to each other 
with constant linear velocity, but not with any rotational motion. Indeed, in a 
rotating frame one is subject to pseudo-forces such as centrifugal forces, which 
disqualifies them from being inertial frames. 

\section{Velocity dependence of inertial-frame relations \label{s3}}

Having established that inertial frames can be related only by constant rotations,
and translations involving uniform motion, we now proceed to determine how these 
tranformation depend on the relative velocity. As by a constant rotation we can 
always align inertial frames to have the same orientation (provided we restrict them
to either all right-handed or left-handed ones, excluding reflections) and we can 
shift their spatial origins to coincide at times $t = t' = 0$ so as to imply $x^a_0= 0$, 
the transformations (\ref{e2.3}) reduce to the explicit form 
\be
t' = \gam t + \kg x, \hs{2} x' = \bg t + \sg x, \hs{2} y' = y, \hs{2} z' = z,
\label{e3.1}
\ee
where $(\gam, \kg, \bg, \sg)$ are constants, which may stil depend on the 
relative velocity $\bfv = (v, 0 ,0)$. As both frames use the same units of length 
and time in laying out their respective co-ordinates, and we exclude reflections 
or time reversal, this transformation must have unit determinant: 
\be
\gam \sg - \kg \bg = 1.
\label{e3.2}
\ee
Now suppose a particle moves in frame $I$ with a velocity ${\bf w} = d{\bfx}/dt$. Then 
its velocity in the frame $I'$, using equation (\ref{e2.2}), is
\be
w^{\prime\,x} = \frac{\bg + \sg w^x}{\gam + \kg w^x}, \hs{1} w^{\prime\, y} = \frac{w^y}{\gam + \kg w^x}, 
\hs{1} w^{\prime\,z} = \frac{w^z}{\gam + \kg w^x},
\label{e3.3}
\ee
and by equation (\ref{e3.2}) the inverse takes the form 
\be
t = \sg t' - \kg x', \hs{2} x = - \bg t' + \gam x', \hs{2} y = y', \hs{2} z = z',
\label{e3.4}
\ee
such that 
\be
w^x = \frac{- \bg + \gam w^{\prime x}}{\sg - \kg w^{\prime x}}, \hs{1} 
w^y = \frac{w^{\prime y}}{\sg - \kg w^{\prime x}}, \hs{1} w^z = \frac{w^{\prime z}}{\sg - \kg w^{\prime x}}.
\label{e3.5}
\ee
In particular, a particle at rest in $I'$: ${\bf w}' = 0$, moves in $I$ with a velocity ${\bf w} = (v, 0, 0)$ 
where
\be
v = - \frac{\bg}{\sg}.
\label{e3.6}
\ee
Necessarily a particle at rest in $I$: ${\bf w} = 0$, then moves in $I'$ with the velocity 
${\bf w}' = (- v, 0, 0)$ where 
\be
- v = \frac{\bg}{\gam}. 
\label{e3.7}
\ee
Therefore the co-efficients of the transformations (\ref{e3.1}) and (\ref{e3.4}) have to satisfy the 
relations
\be
\sg = \gam, \hs{2} \bg = - \gam v. 
\label{e3.8}
\ee
It then follows that 
\be
x' = \gam (x - vt), \hs{2} x = \gam (x' + vt'),
\label{e3.9}
\ee
related by reversing the velocity, as one might have anticipated. 

The final information we need to completely fix the co-efficients of the transformation 
is to require the group property of transitivity of the transformations \ct{poincare:1905}: 
if a third frame $I^{\prime\prime}$, similarly oriented, moves with respect to $I'$ with 
velocity $(v', 0, 0)$, and its co-ordinates are related to those of $I'$ by a transformation
\be
t^{\prime\prime} = \gam' t' + \kg' x', \hs{2} x^{\prime\prime} = \gam' \lh x' - v' t' \rh, \hs{2} 
y^{\prime\prime} = y', \hs{2} z^{\prime\prime} = z',
\label{e3.10}
\ee
then they can be expressed in terms of the co-ordinates of $I$ by
\be \ba{l}
t^{\prime\prime} = \gam^{\prime\prime} t + \kg^{\prime\prime} x 
 = \gam' (\gam t + \kg x) + \kg' \gam (x - vt), \\
 \\
x^{\prime\prime} = \gam^{\prime\prime} \lh x - v^{\prime\prime} t \rh 
 = \gam' (\gam(x - vt) - v' (\gam t + \kg x)),
\ea
\label{e3.11}
\ee
and of course $y^{\prime\prime} = y' = y$ and $z^{\prime\prime} = z' = z$.  
It directly follows that 
\be
\gam^{\prime\prime} = \gam (\gam' - \kg' v) = \gam' ( \gam - \kg v'), \hs{2} 
\kg^{\prime\prime} = \gam' \kg + \gam \kg', \hs{2} 
\gam^{\prime\prime} v^{\prime\prime} = \gam' \gam (v + v'). 
\label{e3.12}
\ee
Now if $v = 0$, clearly the two frames $I$ and $I'$ coincide, and $\gam = 1$, 
$\kg = 0$; similarly if $v' = 0$ then $I'$ and $I^{\prime\prime}$ coincide and
$\gam' = 1$, $\kg' = 0$. However, in the non-trivial case $v,v' \neq 0$ and 
observing that $\gam, \gam' \neq 0$ because of the condition on the 
determinant (\ref{e3.2}), the first equation (\ref{e3.12}) implies that 
\be
\frac{\kg}{\gam v} = \frac{\kg'}{\gam' v'} = - a,
\label{e3.13}
\ee
a {\em universal} constant, as the transformation $I \rightarrow I'$ is completely 
independent of the transformation $I' \rightarrow I^{\prime\prime}$, and therefore 
the velocities $v$ and $v'$ can be chosen at will. Inserting this result back into 
the first equation (\ref{e3.12}): $\gam^{\prime\prime} = \gam \gam' (1 + a v v')$,
the last equation (\ref{e3.12}) gives an equation for the composition of the 
velocities: 
\be
v^{\prime\prime} = \frac{v + v'}{1 + a v v'}.
\label{e3.14}
\ee
Finally, using these results we find that 
\be
\gam^{\prime\prime\,2} \lh 1 - a v^{\prime\prime\,2} \rh = \left[ \gam^2 \lh 1 - a v^2 \rh \right]
\left[ \gam^{\prime\,2} \lh 1 - a v^{\prime\,2} \rh \right],
\label{e3.15}
\ee
and since the velocities $v,v'$ can be chosen freely it follows that the transformation 
$I \rightarrow I'$ between any two inertial frames requires
\be
\gam = \frac{1}{\sqrt{1 - a v^2}},
\label{e3.16}
\ee
where we have chosen the positive square root to guarantee that the times in the 
two frames run in the same direction. Combining the results, the complete most 
general form of the transformation between two equally oriented inertial frames is
\be
t' = \frac{t - a v x}{\sqrt{1 - a v^2}}, \hs{2} x ' = \frac{x - vt}{\sqrt{1 - a v^2}}, \hs{2} 
y' = y, \hs{2} z' = z.
\label{e3.17}
\ee

\section{Discussion \label{s4}}

From the transformation rules (\ref{e3.17}) it is easy to verify that the space-time 
interval
\be
(\Del \tau)^2 = (\Del t)^2 - a \left[ (\Del x)^2 + (\Del y)^2 + (\Del z)^2 \right]
\label{e4.0}
\ee
is invariant under these transformations. Here $\Del \tau$ is the proper time  
interval, measured by a clock in the rest frame. Clearly there are three different 
cases to consider, characterized by the value of the real universal constant 
$a$, which can vanish, can be positive or can be negative. In the simplest 
case $a = 0$ the transformations (\ref{e3.17}) reduce to those of classical 
galilean relativity. In that case time intervals are always equal in all inertial 
frames. This certainly qualifies as a mathematically consistent solution, but 
for all we know not one that describes the correct physics. 

The next case: $a > 0$, allows us to write 
\be
a = \frac{1}{c^2}, 
\label{e4.1}
\ee
where $c$ has the dimensions of a velocity; as $a$ is a universal constant, 
so must be $c$. Indeed, in this case we actually reproduce the standard Lorentz 
transformations, and the constant $c$ can be identified with the speed of light; 
however, note that this identification is not one made {\em a priori}, but it is an 
experimental result that the speed of light appears to be the same in all inertial 
reference frames \ct{michelson:18}. Indeed, it is easily verified that if the first 
velocity $v = c$ in equation (\ref{e3.14}), then also $v^{\prime\prime} = c$
independent of the velocity $v'$. Note furthermore the well-known consequence 
of equation (\ref{e3.14}) that composing velocities less than $c$ one can never 
equal or exceed the value $c$: if $(v, v') < c$ then necessarily $v^{\prime\prime} < c$.
Also the Lorentz transformations imply 
\be 
\Del t = \frac{\Del \tau}{\sqrt{1 - v^2/c^2}},
\label{e4.2}
\ee
and there is a time dilation between clocks in moving frames and in the rest frame. 

Finally, if $a < 0$ we can similarly write 
\be
a = - \frac{1}{c^2},
\label{e4.3}
\ee
where $c$ is a universal constant with the dimensions of velocity. In this case 
the line element (\ref{e4.0}) is that of 4-dimensional euclidean geometry; more 
specifically, it implies that the four-velocity always lies on a fixed 4-dimensional
sphere: if for $a = (1,2,3,4)$ we define $x_a = (x,y,z,ct)$, then
\be
u_a = \frac{dx_a}{d\tau} \hs{1} \mbox{satisfies} \hs{1} \sum_a u_a^{\,2} = c^2.
\label{e4.4}
\ee
Therefore no four-velocity component can exceed $c$. However, the relation 
between four-velocity and co-ordinate velocity $v_i = dx_i/dt$,  is 
\be
\frac{u_i}{c} = \frac{v_i/c}{\sqrt{1 + v^2/c^2}}, \hs{2} 
\frac{u_4}{c} = \frac{dt}{d\tau} = \frac{1}{\sqrt{1 + v^2/c^2}},
\label{e4.5}
\ee
and $v_i$ is not restricted unlike the lorentzian case, although $u_i$ is. Moreover 
we see, that instead of time dilation there is a time contraction between moving frames.
Observe that eq.(\ref{e4.5}) can be inverted, expressing $v_i$ in terms of $u_i$:
\begin{eqnarray}
\frac{v_i}{c} = \frac{u_i/c}{\sqrt{1 - u^2/c^2}} 
\end{eqnarray}
It can be seen that in this case the role of the proper velocity and the relative velocity are reversed compared to their relation in the $a > 0$ case.

As $a$ is a universal constant of nature, it can only have a single  
value. For deciding between the three cases, if one would not be aware 
of the frame-independence of the speed of light already\footnote{For an 
account of the historical development of special relativity, see \ct{pais:1982} 
and references therein.}, it would be most natural to use the comparison of 
proper time and co-ordinate time; in particular, the fact that time-dilation is 
experimentally observed, for example in the life-time of unstable high-energy
particles like muons in cosmic ray showers \ct{gaisser:2016}, decides in the 
favor of positive $a = 1/c^2 > 0$, with the Lorentz transformations providing 
the actual relations between inertial frames. 

The attentive reader may have noticed that the space-time interval (\ref{e4.0}) 
is similar to that of a homogeneous and isotropic universe, with the difference 
that by requiring the universality if clocks and rulers here $a$ is 
fixed to a universal constant, whereas in the cosmological setting the scale factor can be a 
function of cosmic time. However, in the context of standard general-relativistic 
cosmology this is not interpreted as a time-dependent speed of light, but as a 
mismatch between physical distances and co-ordinate differences 
$(\Del x, \Del y, \Del z)$. 

In view of the role and interpretation of the constant $a$, which is more 
general than just a rewriting of Lorentz transformations, we consider $a$ 
more fundamental than the speed of light $c$ itself. Because of the central 
role of the relation (\ref{e4.0}) we propose to call it {\em Minkowski's constant} 
\ct{minkowski:1908}.

\np

\end{document}